\documentclass{ws-procs10x7}

\begin{document}

\title{Dark Matter Searches}

\author{Laura Baudis}

\address{Physics Department, University of Florida, Gainesville, FL 32611, USA\\
E-mail: lbaudis@ufl.edu}

\twocolumn[\maketitle\abstract
{More than 90\% of matter in the Universe could be composed of heavy  
particles, which were non-relativistic, or 'cold',
when they froze-out from the primordial soup. I will review current  
searches for these hypothetical particles,
both via interactions with nuclei in deep underground  
detectors, and via the observation of their annihilation
products in the Sun, galactic halo and galactic center.
 }]

\section{Introduction}

Seventy two years after Zwicky's first accounts of dark matter in galaxy clusters, 
and thirty five years after Rubin's measurements of rotational velocities of spirals,
the case for non-baryonic dark matter remains convincing. Precision observations of the 
cosmic microwave background  and of large scale structures  confirm the picture in 
which more than 90\% of the matter in the universe is revealed only by its gravitational interaction.
The nature of this matter is not known. A class of generic candidates are weakly interacting massive 
particles (WIMPs) which could have been thermally produced in the very early universe. 
It is well known that if the mass and cross section of these particles is determined by the weak scale,
the freeze-out relic density is around the observed value, $\Omega\sim$0.1.
The prototype WIMP candidate is the neutralino, or the lightest supersymmetric particle, which 
is stable in supersymmetric models where R-parity is conserved.  Another recently discussed candidate 
is the lightest Kaluza-Klein excitation (LKP) in theories with universal extra dimensions. If a new 
discrete symmetry, called KK-parity is conserved, and if the KK particle masses are related 
to the weak scale, the LKP is stable and makes an excellent dark matter candidate.
A vast experimental effort to detect WIMPs is underway.  
Cryogenic direct detection experiments are for the first time probing the parameter space predicted by SUSY theories 
for neutralinos, while indirect detection experiments may start to probe the distribution of dark matter 
in the halo and galactic center. In the following, I will give a brief overview of the main search techniques, 
focusing on most recent results. 

\section{Direct Detection}

\begin{figure}[t]
\epsfxsize140pt
\figurebox{140pt}{160pt}{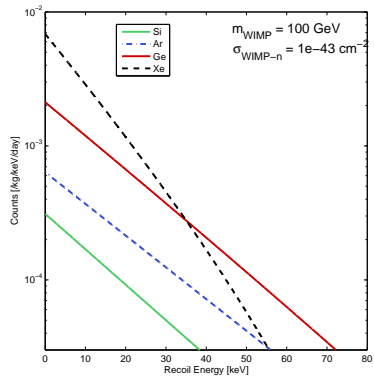}
\caption[]{Differential WIMP recoil spectrum for a WIMP mass of 100\,GeV 
and a WIMP-nucleon cross section $\sigma = 10^{-43}$cm$^2$. The spectrum was 
calculated for illustrative nuclei such as Si (light solid), Ar (light dot-dashed), Ge (dark solid), Xe (dark dashed).} 
\label{rates}
\end{figure}

WIMPs can be detected directly, via their interactions with nuclei in ultra-low-background terrestrial targets\cite{goodman85}.
Direct detection experiments attempt to measure the tiny energy deposition ($<$50\,keV) when a WIMP scatters 
off a nucleus in the target material. Predicted event rates for neutralinos range from 10$^{-6}$ to 10
events per kilogram detector material and day, assuming a typical halo density of 0.3 GeV/cm$^3$. 
The nuclear recoil spectrum is featureless, but depends on the WIMP and target nucleus mass. 
Figure \ref{rates} shows differential spectra for Si, Ar, Ge and Xe, 
calculated for a WIMP mass of 100\,GeV, a WIMP-nucleon cross section of 
$\sigma = 10^{-43}$ cm$^2$ and using standard halo parameters. 

Basic requirements for direct detection detectors are low energy thresholds, low backgrounds and high masses.
The recoil energy of the scattered nucleus is transformed into
a measurable signal, such as charge, scintillation light or phonons, 
and at least one of the above quantities is detected. Observing two  
signals simultaneously yields a powerful discrimination against background events,  
which are mostly interactions with electrons, as opposed to WIMPs and neutrons scattering 
off nuclei. 

In order to convincingly detect a WIMP signal, a specific signature from a particle populating  
our galactic halo is important.
The Earth's motion through the galaxy induces both a seasonal variation of 
the total event rate\cite{fre86,fre88} and a forward-backward 
asymmetry in a directional signal\cite{spergel88,copi99}. 
The expected seasonal modulation effect is  
of the order of $\mathcal{O}$(15kms$^{-1}$/220kms$^{-1}$) $\approx $0.07, requiring
large masses and long counting times as well as an excellent long-term stability 
of the experiment.
The forward-backward asymmetry yields a larger effect, of the order of 
$\mathcal{O}$(v$_\odot$/220km/s)$\approx$ 1, and fewer events are needed 
to discover a WIMP signal\cite{copi99}.
The challenge is to build massive detectors capable of detecting the direction of the 
incoming WIMP. 

\subsection{Experiments}

First limits on WIMP-nucleon cross sections were derived about twenty years ago, 
from at that time already existing germanium double beta decay experiments\cite{ge_exp}. 
With low intrinsic backgrounds and already operating in underground 
laboratories, these detectors were essential in ruling out first WIMP candidates such as 
a heavy Dirac neutrino\cite{beck94}. Present Ge ionization experiments dedicated to 
dark matter searches such as HDMS\cite{hdms} are limited in their sensitivity by 
irreducible electromagnetic backgrounds close to the crystals or  
from production of radioactive isotopes in the crystals by cosmic ray induced spallation.
Next generation projects based on high-purity germanium (HPGe) ionization detectors, 
such as the proposed GENIUS\cite{genius}, GERDA\cite{gerda}, and Majorana\cite{majorana} experiments, 
aim at an absolute background reduction by more than three orders in magnitude, compensating for 
their inability to differentiate between electron- and nuclear recoils on an event-by-event basis. 
Solid scintillators operated at room temperatures had soon caught up with HPGe experiments, 
despite their higher radioactive backgrounds. Being intrinsically fast, these experiments 
can discern on a statistical basis between electron and nuclear recoils, by using the timing 
parameters of the pulse shape of a signal. Typical examples are NaI experiments 
such as DAMA\cite{dama} and NAIAD\cite{naiad}, with DAMA reporting first evidence for a positive 
WIMP signal in 1997\cite{dama-wimp}. The DAMA results have not been confirmed by three different 
mK cryogenic experiments (CDMS\cite{cdms-04}, CRESST\cite{cresst04} and EDELWEISS\cite{edelw05}) and 
one liquid xenon experiment (ZEPLIN\cite{bdmc-04}), independent of the halo model assumed\cite{halo-dama} 
or whether the WIMP-nucleon interaction is taken as purely spin-dependent\cite{spin-dama,savage04}. 
The DAMA collaboration has installed a new, 250\,kg NaI experiment (LIBRA) in the Gran Sasso Laboratory,   
and began taking data in March 2003. With lower backgrounds and increased statistics,  
LIBRA should soon be able to confirm the annual modulation signal. 
The Zaragosa group plans to operate a 107\,kg NaI array (ANAIS) at the Canfranc 
Underground Laboratory (2450 mwe) in Spain\cite{zaragosa-NaI}, and deliver an independent check of the 
DAMA signal in NaI. 
Cryogenic experiments operated at sub-Kelvin temperatures are now leading the field with sensitivities 
of one order of magnitude above the best solid scintillator experiments. Specifically, 
the CDMS experiment can probe WIMP-nucleon cross sections as low as 10$^{-43}$cm$^2$~\cite{prl-si-05}. 
Liquid noble element detectors are rapidly evolving, and seem a very promising avenue towards 
the goal of constructing ton-scale or even multi-ton WIMP detectors.
Many other interesting WIMP search techniques have been deployed, yet it is not the scope of this paper 
to deliver a full overwiev (for two recent reviews see\cite{gabriel05,rick04}).

\subsection{Cryogenic Detectors at mK Temperatures}

Cryogenic calorimeters are meeting crucial characteristics of a successful WIMP 
detector: low energy threshold ($<$10\,keV), excellent energy resolution ($<$1\% at 10\,keV)  
and the ability to differentiate nuclear from electron recoils on an event-by-event basis. 
Their development was driven by the exciting possibility of doing a calorimetric energy 
measurement down to very low energies with unsurpassed energy resolution. 
Because of the T$^3$ dependence of the heat capacity of a dielectric crystal, 
at low temperatures a small energy deposition can significantly change the temperature 
of the absorber. The change in temperature is measured either after the phonons 
(or lattice vibration quanta) reach equilibrium, or thermalize, or when they are still 
out of equilibrium, or athermal, the latter providing additional information 
about the location of an event.

\begin{figure}[t]
\epsfxsize160pt
\figurebox{160pt}{180pt}{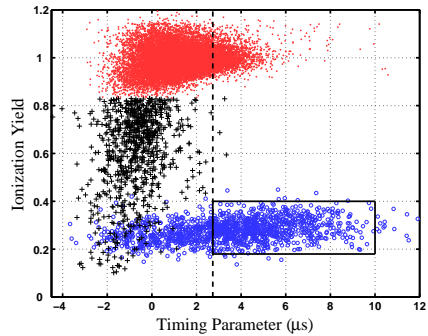}
\caption[]{Ionization yield versus phonon timing parameter for $^{133}$Ba gamma calibration events (dots and crosses)
and $^{252}$Cf neutron calibration events (circles). Low-yield $^{133}$Ba events (crosses) 
have small values of the timing parameter, and the dashed vertical line indicates a timing 
cut, resulting in a high rate of nuclear recoil efficiency and a low rate of misidentified surface events.} 
\label{starttimes}
\end{figure}

\noindent
{\bf CDMS}: the Cold Dark Matter Search experiment operates low-temperature Ge and Si detectors at the 
Soudan Underground Laboratory in Minnesota (at a depth of 2080 m.w.e.).  The high-purity 
Ge and Si crystals are 1\,cm thick and 7.6\,cm in diameter, and have a mass of 250\,g and 
100\,g, respectively. Superconducting 
transition edge sensors photolitographically patterned onto one of the crystal surfaces 
detect the athermal phonons from particle interactions. The phonon sensors are divided into 
4 different channels, allowing to reconstruct the x-y position of an event with a resolution 
of $\sim$1\,mm. If an event occurs close to the detector's surface, the phonon signal is faster 
than for events far from the surface, because of phonon interactions in the thin metallic films.
The risetime of the phonon pulses, as well as the time difference between the charge and phonon 
signals allow to reject surface events caused by electron recoils. Figure~\ref{starttimes} 
shows the ionization yield (ratio of ionization to recoil energy) versus the sum of above timing 
parameters 
for electron recoil events (collected with a $^{133}$Ba source) and nuclear recoil events 
(collected with a $^{252}$Cf source). Events below a yield around 0.75 typically occur within 
0-30\,$\mu$m of the surface, and can be effectively discriminated  while preserving a large part of the nuclear recoil signal.

Charge electrodes are used for the ionization measurement. They are divided into an inner disk, 
covering 85\% of the surface, and an outer  ring, which is used to reject events near the 
edges of the crystal, where background interactions are more likely to occur. 
The discrimination against the electron recoil background is based on the fact that 
nuclear recoils (caused by WIMPs or neutrons) produce fewer charge pairs than electron 
recoils of the same energy. The ionization yield is about 0.3 in Ge, and 0.25 in Si for 
recoil energies above 20\,keV. Electron 
recoils with complete charge collection show an ionization yield of $\approx$1.  
For recoil energies above 10\,keV, bulk electron recoils are rejected with $>$99.9\% efficiency,
and surface events are rejected with $>$95\% efficiency.
The two different materials are used to distinguish between WIMP and neutron interactions by 
comparing the rate and the spectrum shape of nuclear recoil events.

\begin{figure}[t]
\epsfxsize160pt
\figurebox{160pt}{180pt}{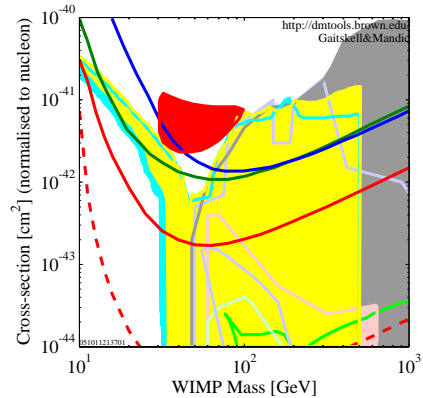}
\caption[]{Experimental results and theoretical predictions for spin-independent WIMP nucleon cross sections 
versus WIMP mass. The data (from high to low cross sections) show the DAMA allowed region (red)\cite{dama-wimp}, 
the latest EDELWEISS result (blue)\cite{edelw05}, the ZEPLIN\,I result (green)\cite{bdmc-04} 
and the CDMS results from 2 towers at Soudan (red)\cite{prl-si-05}. Also shown is the expectation for 5 CDMS towers at Soudan 
(red dashed). The SUSY theory regions are shown as filled regions or contour lines, and are taken from~\cite{susy-theory}. }
\label{limits}
\end{figure}

A stack of six Ge or Si detectors together with the corresponding cold electronics is named a 'tower'.
Five towers are currently installed in the 'cold volume' at Soudan, shielded by about 3\,mm of Cu, 
22.5\,cm of Pb, 50\,cm of polyethylene and by a 5\,cm thick plastic scintillator detector  
which identifies interactions caused by cosmic rays penetrating the Soudan rock.
In 2004, two towers  were operated for 74.5 live days at Soudan, yielding an exposure 
of 34\,kg\,d in Ge and 12\,kg\,d in Si in the 10-100\,keV nuclear recoil energy range. 
One candidate nuclear recoil event at 10.5 keV was observed in Ge, while no 
events were seen in the Si data\cite{prl-si-05}. This result was consistent with the expected background 
from surface events, and resulted in a new upper limit on spin-independent WIMP-nucleon 
cross sections in Ge of 1.6$\times$10$^{-43}$cm$^2$ at the 90\%CL at a WIMP mass of 60\,GeV/c$^2$ 
(see Fig. \ref{limits}).

\begin{figure}[t]
\epsfxsize160pt
\figurebox{160pt}{180pt}{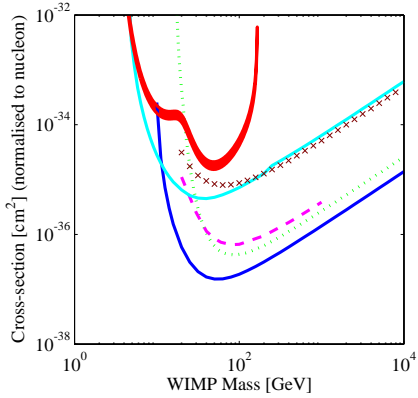}
\caption[]{Experimental results for spin-dependent WIMP couplings (90\% C.L. contours), 
for the case of a pure neutron coupling. The curves (from high to low cross sections) 
show the DAMA annual modulation signal (filled red region), the CDMS Soudan Si data 
(red crosses), the CDMS Stanford Si data (cyan),  EDELWEISS (magenta dashed), 
DAMA/Xe (green dotted) and the CDMS Soudan Ge data (solid blue).} 
\label{limits-cdms-sd}
\end{figure}

The limits on spin-dependent WIMP interactions, shown in  Fig.~\ref{limits-cdms-sd}, are competitive with other experiments, 
in spite of the low abundance of $^{73}$Ge (7.8\%) in natural germanium. In particular, in the case 
of a pure neutron coupling, CDMS yields the most stringent limit obtained so far, 
thus strongly constraining interpretations of the DAMA signal region\cite{savage04,prl-sd-05}. 

\noindent
{\bf EDELWEISS}: the EDELWEISS experiment operates Ge bolometers at 17\,mK in the Laboratoire 
Souterrain de Modane, at  4800\,m.w.e. The detectors are further shielded by 30\,cm 
of paraffin, 15\,cm of Pb and 10\,cm of Cu. They simultaneously detect 
the phonon and the ionization signals, allowing a discrimination against bulk electron 
recoils of better than 99.9\% above 15\,keV recoil energy. The charge signal is measured by Al 
electrodes sputtered on each side of the crystals, the phonon signal by a neutron transmutation 
doped (NTD) heat sensor glued onto one of the charge collection electrodes. The NTD sensors read out 
the thermal phonon signal on a time scale of about 100\,ms. 

\begin{figure}[thbp]
\epsfxsize170pt
\figurebox{170pt}{180pt}{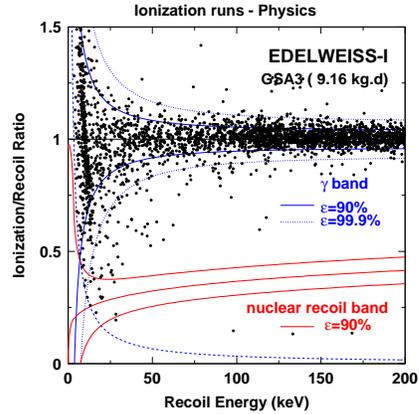}
\caption[]{Ionization yield versus recoil energy for one EDELWEISS 320\,g Ge detector with an 
exposure of 9.16\,kg\,days. Also shown are the electron recoil (blue) and neutron recoil (red) bands.
Figure taken from\cite{edelw05}.} 
\label{edelw-yplot}
\end{figure}

Between 2000-2003, EDELWEISS performed four physics runs with five 320\,g Ge crystals, accumulating 
a total exposure of 62\,kg\,days\cite{edelw05}. Above an analysis threshold of 20\,keV, 
a total of 23 events compatible with nuclear recoils have been observed. Figure \ref{edelw-yplot}
shows the ionization yield versus recoil energy for one EDELWEISS detector for an exposure of 
9.16\,kg\,days. 
The derived upper limits on WIMP-nucleon couplings under the hypothesis that all above events are caused
by WIMP interactions, and for a standard isothermal halo, are shown in Fig.~\ref{limits} and Fig.~\ref{limits-cdms-sd}.

The EDELWEISS experiment has ceased running in March 2004, in order to allow the upgrade to a second phase, 
with an aimed sensitivity of 10$^{-44}$cm$^2$. The new 50 liter low-radioactivity cryostat will be able to house 
up to 120 detectors. Because of the inability of slow thermal detectors 
to distinguish between low-yield surface events and nuclear 
recoils and the inherent radioactivity of NTD sensors, the collaboration has been developing a new design 
based on NbSi thin-film sensors. These films, besides providing a lower mass and radioactivity per sensor, 
show a strong difference in the pulse shape, depending on the interaction depth of an event\cite{nader01}.
The EDELWEISS collaboration plans to operate twenty-one 320\,g Ge detectors equipped with NTD sensors, 
and seven 400\,g Ge detectors with NbSi thin-films in the new cryostat starting in 2005.

\noindent
{\bf CRESST}: the CRESST collaboration has developed cryogenic detectors based on CaWO$_4$ crystals, 
which show a higher light yield at low temperatures compared to other scintillating materials.  
The detectors are also equipped with a separate, cryogenic light detector made of a 30$\times$30$\times$0.4\,mm$^3$
silicon wafer, which is mounted close to a flat surface of the CaWO$_4$ crystal. 
The temperature rise in both CaWO$_4$ and light detector is measured with  tungsten superconducting 
phase transition thermometers, kept around 10\,mK, in the middle of their transition between the superconducting 
and normal conducting state. 
A nuclear recoil in the 300\,g CaWO$_4$ detector 
has a different scintillation light yield than an electron recoil of the same energy, allowing to 
discriminate between the two type of events when both the phonon and the light signals are observed. 
The advantage of CaWO$_4$ detectors is their low energy threshold in the phonon signal, and the fact 
that no light yield degradation for surface events has been detected so far. 
However, about 1\% or less of the energy deposited in the CaWO$_4$ is seen as scintillation 
light\cite{cresst04}. Only a few tens of photons are emitted per keV electron recoil, 
a number which is further 
diminished for nuclear recoils, because of the involved quenching factor. 
The quenching factor of oxygen nuclear recoils for scintillation light is around 
13.5\% relative to electron recoils\cite{cresst04}, leading to a rather high effective recoil energy threshold for  
the detection of the light signal. While neutrons will scatter predominantly on oxygen nuclei, 
it is expected that WIMPs will more likely scatter on the heavier calcium and tungsten. 

The most recent CRESST results\cite{cresst04} were obtained by operating two 300\,g CaWO$_4$ detectors at the Gran Sasso 
Underground Laboratory (3800 m.w.e) for two months at the beginning of 2004. 
The total exposure after cuts  was 20.5\,kg\,days. 
A total of 16 events were observed in the 12\,keV - 40\,keV recoil energy region, 
a number which seems consistent with the expected neutron background, since 
the experiment had no neutron shield at that time. 
No phonon-only events (as expected for WIMP recoils on tungsten) were observed between 12\,keV - 40\,keV 
in the module with better resolution in the light channel, yielding a limit on coherent WIMP 
interaction cross sections similar to the one obtained by EDELWEISS.
CRESST has stopped taking data in March 2004, to upgrade with a neutron shield, an active muon veto, 
and a 66-channels SQUID read-out system. It will allow to operate 33 CaWO$_4$ detector modules, providing 
a total of 10\,kg of target material and a final sensitivity of 10$^{-44}$cm$^2$.

\subsection{\label{liquid_noble}Liquid Xenon Detectors }

Liquid xenon (LXe) has excellent properties as a dark matter detector. It has a high density (3\,g/cm$^3$) 
and high atomic number (Z=54, A=131.3), allowing experiments to be compact. The high mass of the Xe nucleus 
is favorable for WIMP scalar interactions provided that a low energy threshold can be achieved (Fig.~\ref{rates} 
shows a comparison with other target nuclei). LXe is an intrinsic scintillator, having high scintillation ($\lambda$ = 178~nm) 
and ionization yields because of its low ionization potential (12.13~eV).
Scintillation in LXe is produced by the formation of excimer states, which are 
bound states of ion-atom systems. If a high electric field ($\sim$1 kV/cm) is applied, ionization electrons can also
be detected, either directly or through the secondary process of proportional scintillation.
The elastic scattering of a WIMP produces a low-energy xenon recoil,  which loses its energy through ionization and scintillation. 
Both signals are quenched when compared to an electron recoil of the same energy, but by different amounts, allowing 
to use the ratio for distinguishing between electron and nuclear recoils. The quenching factors 
depend on the drift field and on the energy of the recoil. 
At zero electric field, the relative scintillation efficiency of nuclear
recoils in LXe was recently measured to be in the range of 0.13-0.23 for Xe
recoil energies of 10 keV-56 keV \cite{elenaprd-05}. 

\noindent
{\bf ZEPLIN}: the Boulby Dark Matter collaboration has been operating a single-phase LXe detector, ZEPLIN\,I, at the Boulby Mine 
($\sim$3000 m.w.e.) during 2001-2002. ZEPLIN\,I had a fiducial mass of 3.2\,kg of liquid xenon, 
viewed by 3 PMTs through silica windows and  inclosed in a 0.93 ton active scintillator veto. 
A total exposure of 293\,kg\,days had been accumulated. With a light yield of 1.5 electrons/keV, the energy 
threshold was at 2\,keV electron recoil (corresponding to 10\,keV nuclear recoil energy for a 
quenching factor of 20\%). A discrimination between electron and nuclear recoils was applied by using  
the difference in the mean time of the corresponding pulses. 
Using this statistical discrimination method, 
a limit on spin-independent WIMP cross sections comparable to CRESST and EDELWEISS has been achieved 
(see Fig.~\ref{limits}).

The collaboration has developed two concepts for dual-phase detectors, ZEPLIN\,II and ZEPLIN\,III.
ZEPLIN\,II will have a 30\,kg fiducial target mass, observed by 7 PMTs. 
ZEPLIN\,III will operate a lower target mass (6\,kg LXe viewed by 31 PMTs)  at a higher field ($>$ 5\,kV/cm). 
Both ZEPLIN\,II and ZEPLIN\,III are now being deployed at the Boulby Mine and are expected to take data by 2006.

\noindent
{\bf XENON}: the XENON collaboration, including groups from US, Italy and Portugal, 
will operate a 10\,kg dual-phase detector in the Gran Sasso Underground 
Laboratory by 2005-2006. At present, a 3\,kg prototype is under operation above ground, 
at the Columbia Nevis Laboratory. The detector is operated at a drift field of 1\,kV/cm, and
both primary and proportional light are detected by an array of seven 2\,inch PMTs operating in 
the cold gas above the liquid. 
The performance of the chamber was tested with gamma ($^{57}$Co), alpha ($^{210}$Pb) and neutron ($^{241}$AmBe) sources.
The depth of an event is reconstructed by looking at the separation in time between the primary and proportional 
scintillation signal. The x-y position is inferred 
with a resolution of 1\,cm from the center of gravity of the proportional light emitted close to the seven PMTs.
The measured ratio of proportional light (S2)  to direct light (S1) for alpha recoils is 0.03 if the corresponding 
ratio for gamma events (electron recoils) is normalized to 1, providing a very clear separation between these type of events. 

\begin{figure}[thbp]
\epsfxsize160pt
\begin{minipage}[c]{0.5\textwidth}%
\centering\includegraphics[width=0.9\textwidth,height=0.8\textwidth]{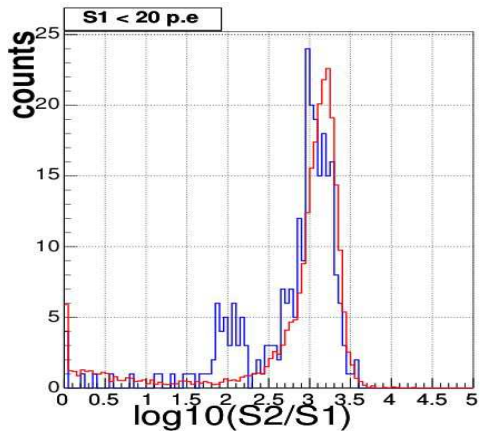}%
\end{minipage}%
\caption[]{Histogram showing the S2/S1 distribution for AmBe events (blue) versus 662\,keV gamma events 
from $^{137}$Cs (red). Two distinct populations are visible in the AmBe data (from\cite{elena-05}).} 
\label{xenon-nrecoils}
\end{figure}

More interesting is the ratio S2/S1 for nuclear recoil events. It was established using a $^{241}$AmBe neutron source, 
by selecting events which were tagged as neutron recoils in a separate neutron detector placed under a scattering 
angle of 130\,deg. If the S2/S1 ratio for electron recoils (provided by a $^{137}$Cs source) is  
normalized to 1, then S2/S1 for nuclear recoils was measured to be around 0.1, the leakage of electron recoils into the 
S2/S1 region for nuclear recoils being $<1\%$ \cite{elena-05}. Figure \ref{xenon-nrecoils} 
shows a histogram of S2/S1 for events taken with the $^{241}$AmBe source, compared to the corresponding distribution 
of events from the $^{137}$Cs gamma source.

The first XENON detector with a fiducial mass of 10\,kg (XENON10) to be operated in Gran Sasso is currently 
under construction. Its goal is to achieve a sensitivity of a factor of 10 below the current CDMS results, 
thus probing WIMP cross sections around 2$\times$10$^{-44}$cm$^2$. 

\subsection{The Future: Direct Detection}

We live in suspenseful times for the field of direct detection: for the first time, a couple of experiments 
operating deep underground probe the most optimistic supersymmetric models. The best limits on WIMP-nucleon 
cross sections come from cryogenic experiments with ultra-low backgrounds and excellent event-by-event 
discrimination power. Although these experiments had started with target masses  
around 1\,kg, upgrades to several kilograms have already taken place or are foreseen for the near future, 
ensuring (along with improved backgrounds) an increase in sensitivity by a factor of 10-100. Other 
techniques, using liquid noble elements such as Xe and Ar, may soon catch up and probe similar parameter spaces 
to low-temperature cryogenic detectors. It is worth emphasizing here that given the importance 
of the endeavor and the challenge in unequivocally identifying and measuring the 
properties of a dark matter particle, it is essential that more than one technique will move forward. 

In supersymmetry, WIMP-nucleon cross sections as low as 10$^{-48}$cm$^2$ are likely\cite{ellis05}. 
Likewise, in theories with universal 
extra dimensions, it is predicted that the lightest Kaluza Klein particle would have a scattering 
cross section with nucleons in the range of 10$^{-46}$ - 10$^{-48}$cm$^2$ \cite{servant-02}. 
Thus, to observe a signal of a few events per year, ton or  multi-ton experiments are inevitable. 
There are several proposals to build larger and improved dark matter detectors. 
The selection presented below is likely biased, but  based 
on technologies which seem the most promising to date. 

The SuperCDMS project\cite{brink05} is a three-phase proposal to 
utilize CDMS-style detectors with target masses growing  from 27\,kg to 145\,kg and up to 1100\,kg, with 
the aim of reaching a final sensitivity of  3$\times$10$^{-46}$cm$^2$ by mid 2015.  This goal 
will be realized by developing improved detectors (for a more precise event reconstruction) and 
analysis techniques, and at the same time by strongly reducing the intrinsic surface contamination 
of the crystals. A possible site is the recently approved SNO-Lab Deep-site facility 
in Canada (at 6000 m.w.e.), where the neutron background would be reduced by more than two orders 
of magnitude compared to the Soudan Mine, thus ensuring the mandatory conditions to build a
zero-background experiment.  In Europe, a similar project to develop a 100\,kg-1\,ton 
cryogenic experiment, EURECA (European Underground Rare Event search with Calorimeter Array) 
\cite{gabriel05} is underway. 
The XENON collaboration is designing a 100\,kg scale dual-phase xenon detector (XENON100), towards 
a modular one tonne experiment\cite{elena-05}.
ZEPLIN MAX, a R\&D project of the Boulby Dark Matter collaboration, 
is a further proposal to build a ton scale liquid xenon experiment. The design will be based on 
the experience and results with ZEPLIN II/III at the Boulby Mine.
WARP\cite{warp} and ArDM\cite{ardm} are two proposals to build ton-scale dark matter 
detectors based on the detection of nuclear recoils in liquid argon. The physics and design 
concepts are similar to the discussed dual-phase xenon detectors. 
The Boulby Collaboration is developing a large directional sensitive detector based 
on the experience with DRIFT\cite{drift}, a time projection chamber  with a total active mass of $\sim$170\,g  
of CS$_2$.

\section{Indirect Detection}

WIMPs can be detected by observing the radiation produced when they annihilate. The 
flux of annihilation products is proportional to $(\rho_{WIMP}/m_{WIMP})^2$, thus regions of interest are 
those expected to have a relatively high WIMP concentration.  
Possible signatures for dark matter annihilation are 
high energy neutrinos from the Sun's core and from the galactic center, gamma-rays from the galactic center and halo 
and antiprotons and positrons from the galactic halo. The predicted fluxes depend on the particle physics model delivering 
the WIMP candidate and on astrophysical quantities such as the dark matter halo profile, the presence of sub-structure 
and the galactic cosmic ray diffusion model.  

\subsection{Gamma Rays}

WIMP annihilation can result in a continuum of gamma rays (via hadronization and decay of $\pi_0$'s) or in a monochromatic 
flux (from direct annihilation into $\gamma\gamma$ or Z$\gamma$), in which case the energy of the gamma line gives direct 
information on the WIMP mass. While the predicted fluxes for the gamma continuum are higher, the energies are lower and 
the signature wouldn't be as clear as in the monochromatic case. In both cases, the expected fluxes are strongly dependent 
on the halo density profile. Direct observation of gamma rays in the energy range of interest to dark matter searches 
(GeV--TeV) can only occur in space, as the gamma will interact with matter via e$^+$e$^-$-pair production, with an 
interaction length much shorter than the thickness of the atmosphere. However, high-energy gamma rays can be detected on the 
ground with air shower detectors. Of these, the atmospheric cerenkov telescopes (ACTs) detect the Cerenkov light produced 
by the cascade of secondary particles in the atmosphere. The background comes from cosmic ray induced showers, 
and imaging ACTs for instance can distinguish between gamma and cosmic ray events based on the light distribution in the Cerenkov cone.
Examples of ACTs either taking data or in construction are HESS\cite{hess}, MAGIC\cite{magic}, 
CANGAROO\cite{cangaroo} and VERITAS\cite{veritas}. 
Their sensitivity typically is in the range 10 GeV - 10 TeV. 
EGRET\cite{egret}, a  space-based telescope on the Compton Gamma Ray Observatory,  took data from 1991-2000 in the energy range 30\,MeV--30\,GeV. 
The next telescope to be launched in 2007 is GLAST\cite{glast} , which will observe the gamma sky up to $\sim$100 GeV. In general, 
space-based telescopes are complementary to ground-based ones, as their range of energies is lower and their field of view and duty cycle larger. 

Recently an excess of high-energy (10$^{12}$ TeV) gammas rays from the galactic center has been detected. This region had been observed by 
the VERITAS and CANGAROO groups, but the angular resolution was greatly increased by the HESS four-telescope array. The HESS 2004 
data confirmed the excess and is consistent with the position of Sgr A$^*$ and a point-like source within the angular resolution 
of the detector (5.8') \cite{horns}. This signal has been 
interpreted as due to WIMP annihilation, with a WIMP mass around 19\,TeV providing the best fit to the data\cite{horns}. 
Apart from the high WIMP mass, the observed signal would require large WIMP annihilation cross sections 
and a cuspy halo profile.

EGRET provided an all-sky gamma-ray survey, with about 60\% of the sources yet to be identified. A reanalysis of EGRET data, 
which is publicly available, revealed that the diffuse component shows an excess by a factor of two above the 
background expected from $\pi^o$'s produced in nuclear interactions, inverse Compton scattering and bremsstrahlung.  
This excess, which is observed in all sky directions, has been interpreted as due to WIMP annihilation, with a best fit WIMP mass around 
60\,GeV~\cite{deboer}. The relative contributions 
of the galactic background have been estimated with the GALPROP code\cite{galprop}, while the extragalactic background was obtained 
by subtracting the galactic 'foregrounds' (as given by GALPROP) from the EGRET data. The predicted cross section for elastic scatters 
on nucleons is in the 5$\times$10$^{-8}$-2$\times$10$^{-7}$pb range, and thus testable by the CDMS experiment. 

An analysis of the EGRET extragalactic background revealed two components, a steep spectrum power law with index $\alpha$=-2.33 and a 
strong bump at a few GeV~\cite{wuerzburg}. Such a multi-GeV bump is difficult to explain in conventional astrophysical models, 
with contributions 
from faint blazars, radio-galaxies, gamma-ray bursts and large scale structures, and has been 
interpreted as evidence for WIMP annihilation\cite{wuerzburg}. The best fit is provided by a WIMP mass around 500\,GeV and 
an annihilation cross section of $\langle\sigma v\rangle \approx 3 \times10^{-25}cm^3s^{-1}$. A typical candidate from supersymmetry 
would have cross sections 
on nucleons $<$10$^{-7}$pb, thus below the present sensitivity of direct detection experiments at these masses.
However, while a $>$500\,GeV WIMP would be difficult to detect at the LHC (and even at the ILC), it is within the reach of 
next generation direct detection experiments. 

\subsection{Neutrinos from the Sun}

WIMPs with orbits passing through the Sun  can scatter from nuclei and lose kinetic energy. If their final 
velocity is smaller than the escape velocity, they will be gravitationally trapped and  will settle 
to the Sun's core. Over the age of the solar system, a sufficiently large number of WIMP can accumulate and efficiently 
annihilate, whereby only neutrinos are able to escape and be observed in terrestrial detectors. 
Typical neutrino energies are 1/3--1/2 of the WIMP mass, thus well above the solar neutrino background. Observation of 
high energy neutrinos  from the direction of the Sun would thus provide a clear signature for dark matter in 
the halo. The annihilation rate  is set by the capture rate, which scales with (m$_{WIMP}$)$^{-1}$ for a given halo 
density. Thus, annihilation and direct detection rates have the same scaling with the WIMP mass. However, the probability of 
detecting a neutrino by searching for muons produced in charge-current interactions scales with E$_{\nu}^2$, making 
these searches more sensitive at high WIMP masses when compared to direct detection experiments. 
The best technique to detect high-energy neutrinos is to observe the upward-going muons produced in charged-current interactions
 in the rock below the detector. To distinguish neutrinos coming from the Sun's 
 core from backgrounds induced by atmospheric neutrinos, directional information is needed. The direction of the upward-going muon 
 and the primary neutrino direction are correlated, the rms angle scaling roughly with $\sim$20$(E_{\nu}/10GeV)^{-1/2}$.

 Two types of detectors are used to search for high-energy neutrino signals,  with no excess above the atmospheric 
 neutrino background  reported so far. 
 In the first category are large underground detectors, such as MACRO\cite{macro} and SuperKamiokande 
\cite{superk}, while 
 the second type are dedicated neutrino telescopes, employing large arrays of PMTs deep in glacier ice, in the ocean or in a lake, 
 such as AMANDA\cite{amanda}, BAIKAL\cite{baikal}, NESTOR\cite{nestor} and ANTARES\cite{antares}. These experiments detect the Cerenkov light emitted 
 when muons move with speeds larger than the velocity of light in water/ice, with $\sim$1\,ns timing resolution. The PMT hit pattern 
 and relative arrival times of the photons are used to reconstruct the direction of the incoming particle, which is correlated with 
 the direction of the neutrino. Neutrino telescopes have higher energy thresholds (in the range 50-100\,GeV), but their effective area 
 is much larger, thus compensating for the lower fluxes predicted for heavy WIMPs. 
 
\begin{figure}
\epsfxsize160pt
\figurebox{160pt}{180pt}{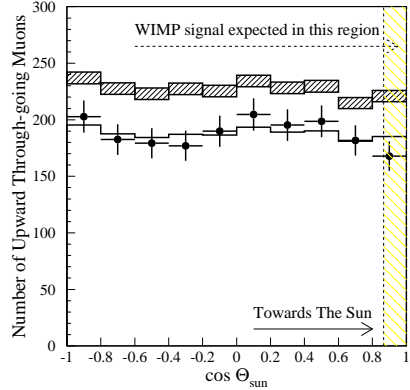}
\caption[]{Angular distribution of upward through-going muons with respect to the Sun in Super-Kamiokande. Data are black circles, 
while the hatched region and the solid line show the expected atmospheric neutrino background before and after taking into 
account $\nu$-oscillations. Figure from\cite{superk}.}
\label{superk-fig}
\end{figure}

\begin{figure}
\epsfxsize160pt
\figurebox{160pt}{180pt}{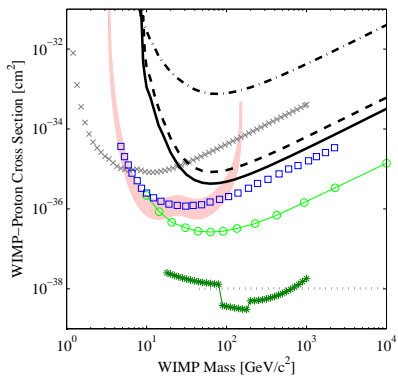}
\caption[]{Upper limits on WIMP-nucleon cross sections for pure proton coupling. The lowest curve (green asterisks) is the 
SuperKamiokande limit, the filled region shows the 3-$\sigma$ DAMA region. For details, see\cite{prl-sd-05}.}
\label{superk-limits}
\end{figure}

 The strongest limits on high-energy neutrinos coming from the Sun are placed by Super-Kamiokande\cite{superk}. 
 Figure  \ref{superk-fig} shows the angular distribution of upward through-going muons with respect to the Sun, 
 and the expected atmospheric neutrino background in the Super-Kamiokande 
 detector. While the limits on scalar WIMP-nucleon interactions are not competitive to direct detection 
 experiments, Super-Kamiokande gives the most stringent limit on spin-dependent WIMP-nucleon cross sections 
 for pure proton couplings above a WIMP  mass of $\sim$20\,GeV  (see Fig. \ref{superk-limits}).

\subsection{Positrons and antiprotons}

Cosmic positrons and antiprotons produced in WIMP annihilations in the galactic halo can be observed 
with balloon or space-based experiments. For antiprotons, the background spectrum produced by spallation 
of primary cosmic rays on H atoms in the interstellar medium is expected to fall steeply at energies below 
1 GeV. Thus, observation of low-energy cosmic-ray antiprotons could provide evidence for WIMPs in the halo. 
The BESS collaboration has measured the flux of antiprotons in several balloon flights between 
1193-1997 \cite{bess}. No convincing excess above the cosmic ray background has been observed. 
For positrons, the background flux  is expected to 
decrease slowly as a function of energy. The HEAT experiment has measured a positron excess at energies $\sim$8\,GeV \cite{heat}, 
which can been interpreted as coming from WIMP annihilation in the halo\cite{edsjo}. The spectra can be fit by a WIMP mass of 
200-300\,GeV, but the fit is far from perfect. The signal requires a boost of a factor of $\sim$30 in the WIMP density,  for instance from 
the presence of dark matter clumps in the halo.

\subsection{Future: Indirect Detection}

Several existing observations have been interpreted as signatures for dark matter annihilation 
in our galaxy, or in extragalactic dark matter halos. There is no single 
WIMP capable of explaining all the data: WIMP masses from 60\,GeV to  18\,TeV are required,
with large boost factors in the halo density and  a cuspy inner halo. 
There is a clear demand for more data. Existing and future ACTs, such as HESS, MAGIC, 
CANGAROO and VERITAS will map out the galactic center with improved position, energy and timing resolution, and 
will likely reveal the source of high-energy gammas. Planned space-based detectors such as GLAST 
will map the gamma-ray sky with unsurpassed sensitivity, while 
cubic kilometer detectors in ice (IceCube) or  water (Km3Net) 
will considerably increase existing sensitivities to high-energy neutrinos coming from the Sun. 
Finally, PAMELA, to be launched in 2007, and AMS-2, to be operated on the ISS starting in 2007 will 
greatly improve the sensitivity to antimatter from WIMP annihilation in the galactic halo.

In looking back over the fantastic progress made in the last couple of years, and extrapolating into the future,
it seems probable that these, and other proposed projects, will have a fair chance to discover a WIMP 
signature within the present decade. 
In conjunction with direct WIMP searches and accelerator production of new particles at the weak scale, 
they will allow to reveal the detailed properties of WIMPs, such as their mass, spin and couplings 
to ordinary matter, and shed light on the density profile of dark matter in the halo.

 \clearpage
\twocolumn[
\section*{DISCUSSION}
]

\begin{description}
\item[Guy Wormser] (LAL Orsay):\\
  Can you translate the limits on results obtained by indirect 
 experiments like HESS on to the parameter space of direct 
 searches?

\item[Laura Baudis:] HESS observes a signal, which can be interpreted as due to 
WIMP annihilation in the galactic center, with a WIMP mass higher than 12\,TeV. 
In general, direct searches have reduced sensitivity  
at such high WIMP masses, because of the lower fluxes for a given halo density. 
The WIMP-nucleon scattering cross section depends on the particle physics candidate. 
For instance, for a scalar cross section larger than $\approx$4$\times$10$^{-42}$cm$^2$, 
a 20\,TeV particle would be ruled out by the CDMS experiment. 

\item[Peter Schuber] (DESY):\\
  You did not mention the recent publication in Phys. Rev. Lett.
from the University of Wurzburg that reported a plausible evidence for 
neutralinos from satellite based high energy photon spectra. In fact 
they indicated even a high neutralino mass of about 500\,GeV.

\item[Laura Baudis:] I misunderstood this question. I did not know about above publication, 
but have included it in these proceedings. A 500\,GeV neutralino, although likely beyond 
the reach of planned accelerators, can be probed by future direct detection experiments.

\item[Bennie Ward] (Baylor University):\\
  On the slide wherein you exhibited your candidate event as one of 
many background events presumably associated with a detector 
mis-function, what happened to the other events that wear near it?

\item[Laura Baudis:] The other events did not survive the standard CDMS cuts 
(such as data quality, fiducial volume, ionization yield and timing cuts). 

\end{description}

\end{document}